# Cavity Plasmon: Enhanced Luminescence Effect on InGaN Light Emitting Diodes

Yuyin Li[1], Jing Zhou[2,*], Ziwen Yan[1], Xianfei Zhang[1], Zili Xie[1,], Xiangqian Xiu[1], Dunjun Chen[1], Bin Liu[1], Hong Zhao[1], Yi Shi[1], Rong Zhang[1], Youdou Zheng[1], and Peng Chen[1,*]

[1]*Jiangsu Provincial Key Laboratory of Advanced Photonic and Electronic Materials and School of Electronic Science and Engineering, Nanjing University, Jiangsu, Nanjing, 210093, China.*
[2]*School of integrated circuits, Anhui University, Anhui, Hefei, 230039, China.*
[*]**Authors to whom correspondence should be addressed:** 22126@ahu.edu.cn, pchen@nju.edu.cn

## Abstract

We fabricated polygonal nanoholes in the top p-GaN layer of the InGaN/GaN light-emitting diode, followed by the deposition of Au/Al metal thin film within the nanoholes to create metal microcavities, thereby constructing the surface plasmon structure. The findings indicate that with increased current injection, the light output of the LEDs rose by 46%, accompanied by a shift of the gain peak position towards the plasmon resonance energy. The maximum enhancement factor increases to 2.38 as the coupling distance decreases from 60 nm to 30 nm. Interestingly, time-resolved photoluminescence data showed that the spontaneous emission decay time lengthened due to the plasmon coupling, suggesting the presence of a new plasmon coupling mechanism. Finite-Difference Time-Domain simulation results show that the electric field is localized at certain locations around the metal microcavity, generating a new type of shape-sensitive plasmon, named **Cavity Plasmon** here. This intense localization leads to a longer lifetime and enhances the recombination efficiency of excitons. We discuss several unique properties of the cavity plasmon generated by the polygonal metal microcavity with several specific angular shapes. The results demonstrate that the cavity plasmon generated by the polygonal metal microcavity is a highly promising technique for enhancing the light emission performance of of relevant semiconductor optoelectronic devices.

**Keywords:** surface plasmon, cavity plasmon, spontaneous emission lifetime, coupling mechanism.

## 1. Introduction

In recent years, GaN-based LEDs have been widely used in solid-state lighting, large-area display, and visible light communication because they can basically replace traditional lighting sources, showing less energy loss, longer life, and higher efficiency. Many research works have been carried out to address the adverse factors restricting the light-emitting efficiency, trying to further improve the light-emitting efficiency. [1,

2]. Among the many techniques to improve the quantum efficiency in InGaN-based LEDs, surface plasmon (SP) coupled enhanced luminescence is a very promising technique. SP is a collective oscillation mode of electrons in a material, which can break through the diffraction limit, couple with the external field strength, and obtain a large near-field enhancement. SP-QW coupling leads to the enhancement of light emission since they were first directly observed experimentally in 2004[3], SP-coupled LEDs have gained widespread attention.

SP can be used for luminescence enhancement of LEDs, mainly due to the strong coupling effect of SP and excitons in MQWs. At present, there are two modes of MQWs-SP coupling:

(1) Metal nanoparticles lead to enhanced spontaneous emission. It is generally believed that the time constant of the surface plasmon excited by the recombination energy is very short, so more energy can be transferred to the SP. Once the radiation efficiency of SP is large enough, an enhancement of spontaneous emission can be observed [4-8]. For instance, *Cho* et al. noticed higher PL and EL intensities after coating the p-GaN of a green LED with a layer of gold nanoparticles[7, 8].

(2) S SP laser based on metal-dielectric coupling. This is a metal-dielectric coupled SP laser, which has been reported in many materials. The most typical is the semiconductor-insulator-metal (SIM) structure such as the GaN triangular nanowire-$SiO_2$ (8nm)-Al film (100nm) system proposed by *Zhang et al.* in 2014[9], and *Sidiropoulos et al.* proposed ZnO cylindrical nanowires-LiF(10nm)-Ag film(100nm) system[10]. When the emitter is coupled with the SP generated by the metal film through the insulating medium, the energy is concentrated in the dielectric layer between the emitter and the metal film, and SP lasing is obtained at a high injection level. By transferring exciton energy to SP, a new recombination channel is created, which can boost spontaneous emission rates and even produce SP lasers. It can be found that the energy transfer in this route is unidirectional, and there is no process of SP energy feedback back to excitons in the whole process.

According to the Purcell effect, the spontaneous emission rate of carriers in the optical cavity will be increased compared to that in a regular air environment. The Purcell factor F is utilized to quantify this enhancement in spontaneous emission, reflecting the overall efficiency of the system to some extent. The strong electric field confined within the light-emitting structure is responsible for inducing the Purcell factor through the surface plasmon mode. This localized electric field plays a crucial role in the interaction between excitons and SP. Fortunately, by manipulating the shape of the metal surface plasma, the oscillation mode of electrons can be controlled, thereby regulating the local electric field of the nanostructure. For example, in triangular structures, the maximum dipole resonance intensity occurs at their sharp corners.[11], The geometry of metal nanostructures, such as spheres, cubes, tetrahedrons, and hexahedrons, significantly impacts the local surface plasmon spectrum [12]. Additionally, the sharpness of corners or edges in nanostructures greatly influences their surface plasmon properties [6, 13, 14]. These resonances and associated strong fields arise from the accumulation of polarization charges on the surface of the plasmon structure[11]. The field distribution around non-spherical nanostructures can be particularly intense near sharp tips, making them suitable for near-field enhancement applications. Therefore, a well-thought-out design of plasmon structures can lead to a remarkable enhancement in field strength.

In this study, we presented a novel plasmonic LED structure utilizing InGaN/GaN LEDs with Au/Al polygonal microcavities. The presence of sharp edges on polygonal metal microcavities can enhance luminescence through increased plasmon effects via a super-strong field enhancement mechanism. Our experimental findings, supported by time-resolved photoluminescence (TRPL) studies, revealed an extended exciton lifetime, indicating a unique plasmon mechanism compared to previous research. We propose a secondary energy transfer pathway, where the Cavity Plasmon (CP) can alter the recombination dynamics of excitons within the evanescent field, leading to improved emission efficiency via either accelerated spontaneous emission or prolonged exciton lifetime in quantum wells (QWs). Finite-difference time-domain (FDTD) simulations further illustrate the close correlation between plasmon properties and the microcavity's geometry. Besides enhancements in the near-evanescent field, polygonal microcavities with sharp corners also exhibit far-field improvements in regions distant from the metal. This localized field enhancement plays a crucial role in enabling the realization of the secondary energy transfer pathway.

## 2. Materials and Methods

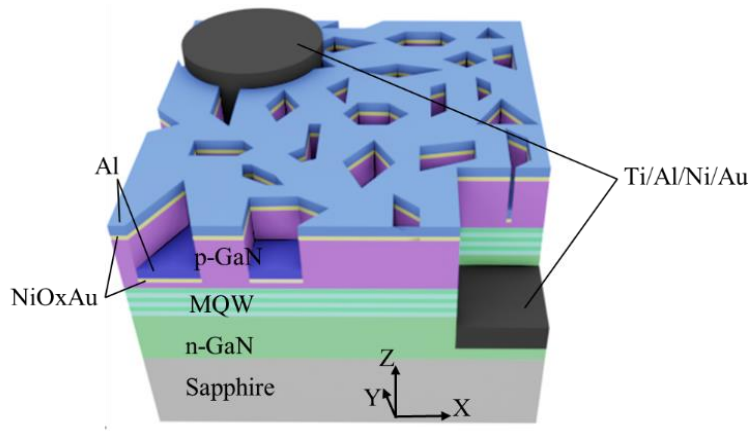

Figure 1. The structure of the LED.

The LED wafer used in this study was grown on a c-plane sapphire substrate by metal organic chemical vapor deposition (MOCVD). The device structure is shown in Figure 1, including 4-μm undoped and Si-doped GaN, five-period InGaN/GaN QWs, and 200-nm p-type GaN. For the fabrication of the polygonal nanoholes, a 6-nm Ni was deposited on the LED surface by electron beam evaporation (EBE), following with rapid thermal annealing (RTA) to form nanoporous mask with proper temperature and duration. Then the LED wafer was etched with the nanoporous mask by inductively coupled plasma etching (ICP). In order to study the influence of different coupling distances, the etching depth was controlled at 140 nm and 170 nm respectively. By this way, LED chips with three structures on the p-type GaN layer are obtained, one is the unetched planar p-GaN (plane-LED), another two are nanoporous p-GaN with a depth of 140 nm (pore140-LED) and with a depth of 170 nm (pore170-LED). After etching, the remaining Ni film was washed away with dilute nitric acid.

Next, Ni/Au (5/5 nm) was deposited on the whole p-GaN layer and annealed at 500°C in air for 1 min. As Ni will be oxidized to $NiO_x$ during annealing, the metal at the bottom of the hole was 5-nm Au only [15]. Then, 20-nm Al was deposited at last, so that the metal at the bottom of the hole is Au/Al (5/20 nm) double layer structure. The last step, Ti/Al/Ni/Au (30/170/50/150 nm) was deposited by electron beam evaporation as n-type electrode and p-type electrode at designated positions.

The p-type GaN surface nanoporous structure was characterized by scanning electron microscope (SEM) and atomic force microscopy (AFM). For LED devices with different etching depths of p-GaN, electroluminescence (EL) tests were performed from the top surface and backside of the LED devices, respectively.

## 3. Results and Discussion

*3.1 Experimental characterization of porous mask and p-GaN surface*

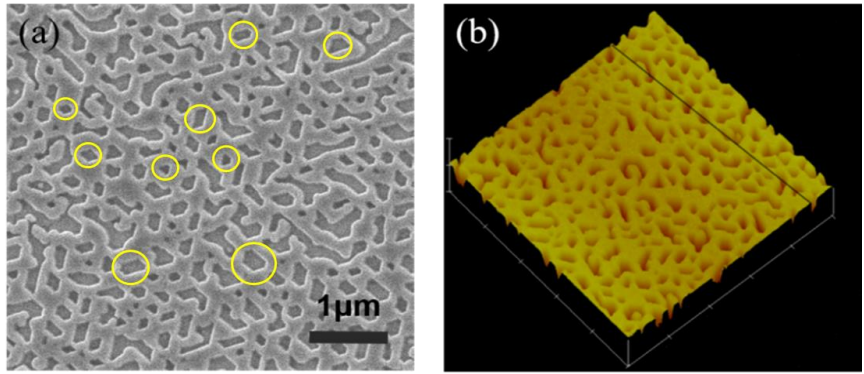

Figure 2. The morphology of the nanoporous structure on the p-type GaN surface: (a) SEM image of the Ni porous template, (b) AFM image of the p-type GaN surface nanoporous structure of pore170-LED

Figure 2 shows the surface morphology of the nanoporous structure realized after annealing Ni on the p-GaN surface. It can be seen from Figure 2 (a) that the nanoholes formed after annealing of the Ni film, which present a polygonal morphology. The polygonal nanoholes have the common feature that the corners are 60° or 120°[16]. Figure 2 (b) is an AFM image of the etched p-GaN surface. After comparison with Figure 2 (a), it can be seen that hole patterns have been successfully transferred from the Ni nanoporous mask to the p-GaN layer. In this sample, the hole depth is basically maintained at 170 nm. Considering that the thickness of p-GaN is 200 nm, the distance between the bottom of the pore170-LED and the active area is only 30 nm, which is less than the penetration depth from nomal SP theory. The penetration depth (Z) can be approximately given by,

$$Z = \lambda/2\pi [\,(\varepsilon'_{GaN} - \varepsilon'_{metal})/\varepsilon'^{2}_{metal}\,]^{1/2} \qquad (1)$$

where $\varepsilon'_{GaN}$ and $\varepsilon'_{metal}$ are the real parts of the dielectric constants of GaN and metal. Based on this equation, Z can be calculated as 77 nm for Al and 33 nm for Au, respectively[3]. Thus, in our samples, Al should give a more important role by considering Al thicker thickness and its Z value.

*3.2 The I-V and Optical Characteristics of the LEDs*

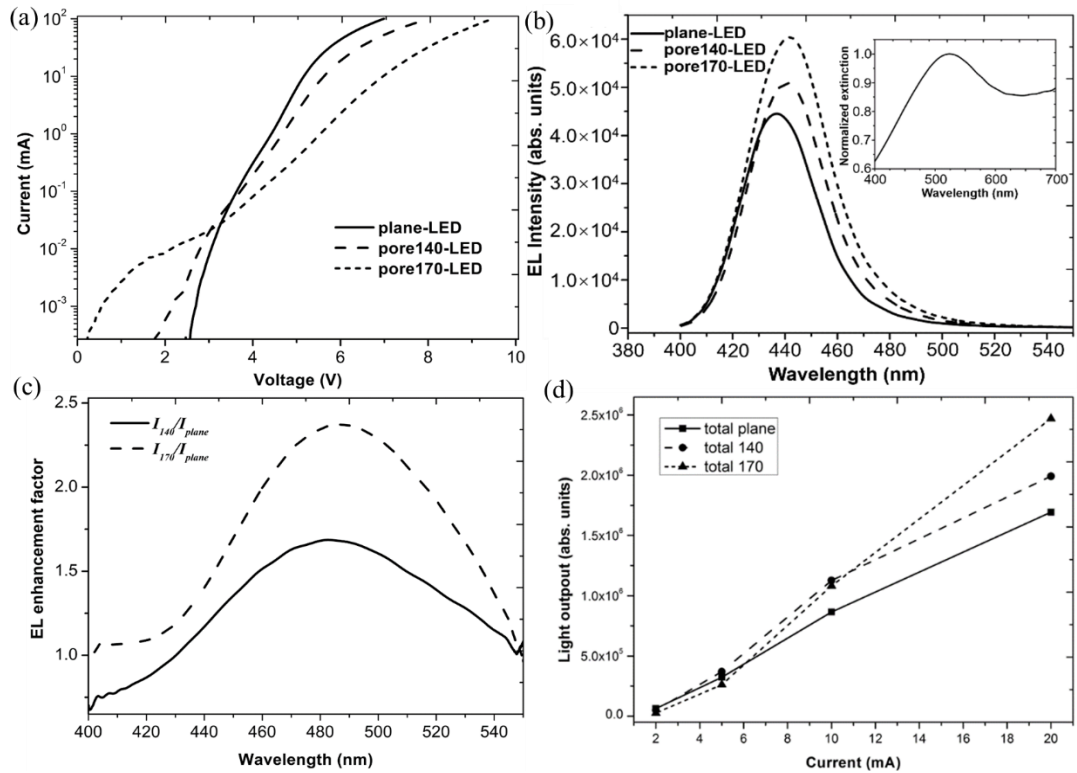

Figure 3. (a) I-V curves of plane-LED, pore140-LED and pore170-LED. (b) The EL spectrum of the total light output power by the sum of the front and back light; the inset is the absorption spectrum of Au/Al particles. (c) EL enhancement factor curve of pore140-LED and pore170-LED. (d) Light output power curve of plane-LED, pore140-LED and pore170-LED under different working currents

Figure 3(a) shows the I-V characteristic curves of the three structures. In the case of low voltage (<3 V), as the depth of the hole increases, the forward currents of the LEDs are larger than the plane LED, which is mainly caused by the leakage current due to the etching damages. When the voltage is above 3 V, as the depth of the hole increases, the forward currents of the LEDs are lower than the plane LED, which is mainly caused by the higher resistance due to the smaller effective injection area of the p electrode.

It has been known that the surface roughness caused by the nanoporous p-type surface can increase the light extraction efficiency of the top surface, which increases the light scattering at the top surface and then reduces the total reflection[17]. At the same time, the surface-plasmon coupling can also increase the emission intensity, but not limited on the top surface[18]. Therefore, it is necessary to perform the EL spectrum test from the top and back side of the LED devices at the same time.

Figure 3 (b) shows the EL spectra of the plane-LED, pore140-LED and pore170-LED at a current of 20 mA, which are already the sum of the top and bottom emission spectra. The inset shows the absorption spectrum of Au/Al particles embedded in the p-GaN hole, and it indicates that the peak of the Au/Al SPR is at 523 nm in the extinction spectrum. From Figure 3(b), a redshift can be seen for the devices with SP coupling, and the redshift is increased with the increase of the hole depth, *i.e.* the stronger coupling. As the hole depth increases, the spectrum peak of the LED gradually shows the redshift from 435.9 nm of plane-LED to 439.9 nm of pore140-LED and 441.6 nm of pore170-LED respectively. The strain relaxation[16] and higher current density[19] will cause the blue shift of the emission peak wavelength, which is contrary to our experimental observation. So, the surface roughing may partially contribute to the increase of the light output, but there are other reasons. On the other hand, by comparing the extinction spectrum with the EL spectrum, it can be found that the EL change satisfies the typical phenomenon of SPs coupling, that is, the peak of the EL spectrum moves to the direction of surface plasmonic resonance (SPR) peak. As metal get closer to the MQWs, the luminescence peak shift amplitude increases, further to the SPR peak[3, 20]. Therefore, it can be concluded that the main reason for the EL spectrum redshift in the coupled LED is the coupling of Au/Al film and the MQWs.

Besides the redshift, the emission intensity has also been modulated by the surface structure. The emission intensity, $I_{plane}$, is for the plane-LED, and similarly $I_{140}$ is for pore140-LED, $I_{170}$ is for pore170-LED. By calculating the ratio between the coupled samples and the plane-LED, we can obtain the wavelength-depended EL enhancement factor curve, $I_{140}/I_{plane}$ and $I_{170}/I_{plane}$, as shown in Figure 3(c). For pore140-LED, the maximum value of EL enhancement factor is 1.69, and the corresponding wavelength is 482.1nm. As the hole depth increases from 140 nm to 170 nm, the maximum enhancement factor increases to 2.38, and the peak wavelength shifts to 486.1 nm. Considering the exponential decay characteristics of the SP evanescent field, the increase of EL enhancement factor and the redshift of the enhancement peak with the increase of the hole depth can be explained by the stronger coupling with the closer distance. As the distance between Au/Al film and MQWs decreases from 60 nm to 30 nm, the QW-SP coupling becomes stronger.

In order to further investigate the properties of LEDs, the relationship between the EL spectrum integrated intensity and the current was tested for all three samples, as shown in Figure 3(d). In the case of low current (I = 2 mA), the integrated EL intensity of the plane-LED is higher than those of the pore140-LED and the pore170-LED. With the increase of current, the integrated EL intensities of the pore170-LED and the pore140-LED is higher than that of the plane-LED gradually. When the current is 20 mA, the integral intensities of the pore140-LED and the pore170-LED are 18% and 46% higher than that of the plane-LED respectively. Considering the above analysis, it can be seen that with the working current of 20 mA, a stronger enhancement can be obtained in the stronger coupled sample. In the case of low current, the integrated intensity of the SP-coupled LEDs is smaller than that of the planar device, which may be due to the large leakage current[21] of the porous structure caused by the lattice damage from the dry etching [22]. However, as the working current increases, the leakage current will reach saturation and the total current mainly comes from the diffusion current. This means that more current can contribute to the recombination in the active area under high current injection, so the effect of SP enhancement is prominent now. It shows a special meaning that the stronger enhancement is observed by the effect of SP coupling at a higher working current. It is well known that the InGaN-based LEDs always encounter a serious problem, i.e. efficiency droop at high current injection. So, from this study, the SP coupling may provide a very promising solution for high-efficiency InGaN-based LEDs with the second energy transfer route.

*3.3 Time-resolved photoluminescence from LEDs.*

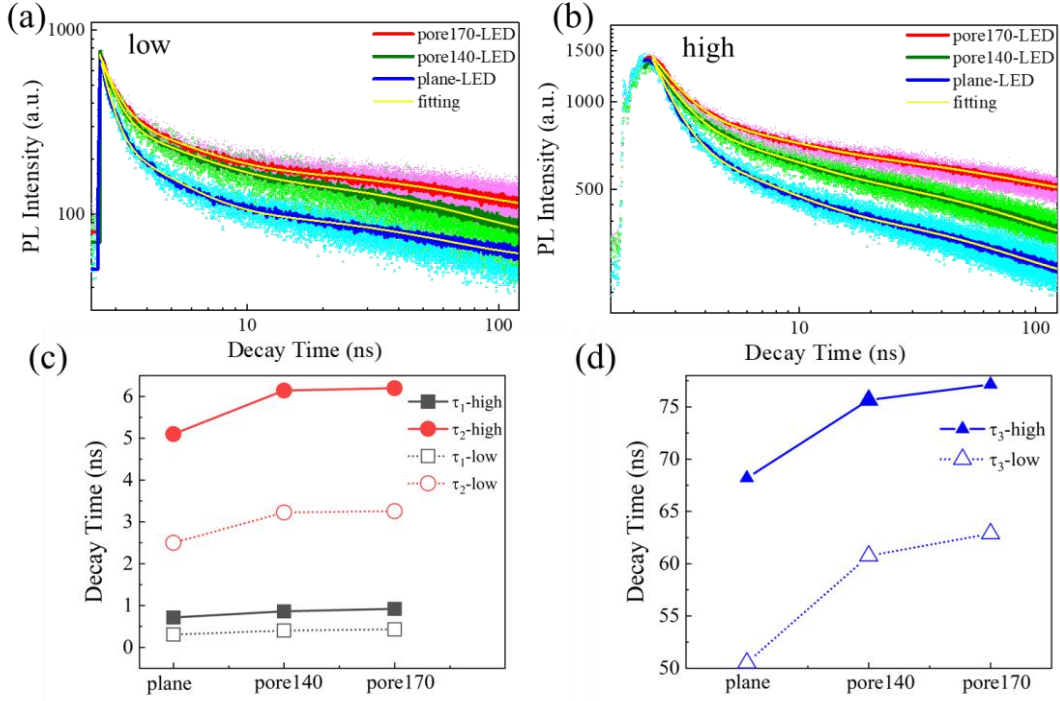

Figure 4. Time-resolved PL (TRPL) measurement results at 300 K under low excitation power density (a), and high excitation power density (b). Decay time $\tau_1$ and $\tau_2$ (c), $\tau_3$ (d).

In order to investigate the plasmon-exciton coupling mechanism. The TRPL experiments were carried out at room temperature under different excitation powers . Typical result was showed in Figures 3a and 3b. A triple exponential function can be used to fit the decay curves of the TRPL, and the carrier lifetime can be determined by fitting the PL decay curves using the following equation:

$$I_t = A_1 e^{-t/\tau_1} + A_2 e^{-t/\tau_2} + A_2 e^{-t/\tau_3} \quad (2)$$

where I(t) is the PL intensity at time t, $A_1$ and $A_2$ and $A_3$ are constants and $\tau_1$、$\tau_2$ and $\tau_3$ correspond to three-process decay time constant, respectively. This three-step PL decay with early fast and later slow decay processes different, it also includes a moderate-rate decay process. Previous studies suggest that SP-MQW coupling affects the fast decay process, and exhibits a much faster decay time[7, 23, 24].

However, our LEDs exhibit longer emission times, and the exponential fitting results are also shown with yellow lines in Figure 4(a) and Figure 4(b), as shown in Figure 4(a) and Figure 4(b). Under a high excitation power density, the $\tau_1$ of the plane-LED, pore140-LED and proe170-LED samples are 0.71, 0.85 and 0.92 nanoseconds (ns). The $\tau_2$ of the plane-LED, pore140-LED and proe170-LED samples are 5.10, 6.14 and 6.20 ns. The $\tau_3$ of the plane-LED, pore140-LED and proe170-LED samples are 68.20, 75.65 and 77.15 ns. The contrast relationship is shown by the solid line in Figure 4(c) and 4(d). The decay time of the three processes is increased compared with that of plane-LED. The same trend is shown at low power densities. This demonstrates that the plasmons structure of the polygonal microcavity in our LEDs affects the entire recombination process of the excitions in our LEDs, which differs from earlier studies. And it can be seen that as the metal and the MQW get closer and closer, the coupling effect gradually increases, as does the exciton lifetime.

## 4. Finite-Difference Time-Domain Simulation Results.

Because the metal sheet is in a bounded cavity, its electron oscillations are not just the ordinary surface oscillations, but also have a strong link with the geometry of the microcavity, which is the primary reason for the varied coupling effects. The FDTD approach is used to determine electric filed energy distribution in the structure in order to study the emission mechanism of the features of the cavity plasmons. A three-dimensional FDTD simulation with all perfectly matched layer (PML) boundaries was used. The hole's depth is H. To be compatible with the experiment, the GaN surface and bottom of the hole are covered with a 5 nm Au and 20 nm Al coating. The refractive index of GaN is n = 2.4. The dielectric constants of Au and Al are taken from the data described Palik[25].

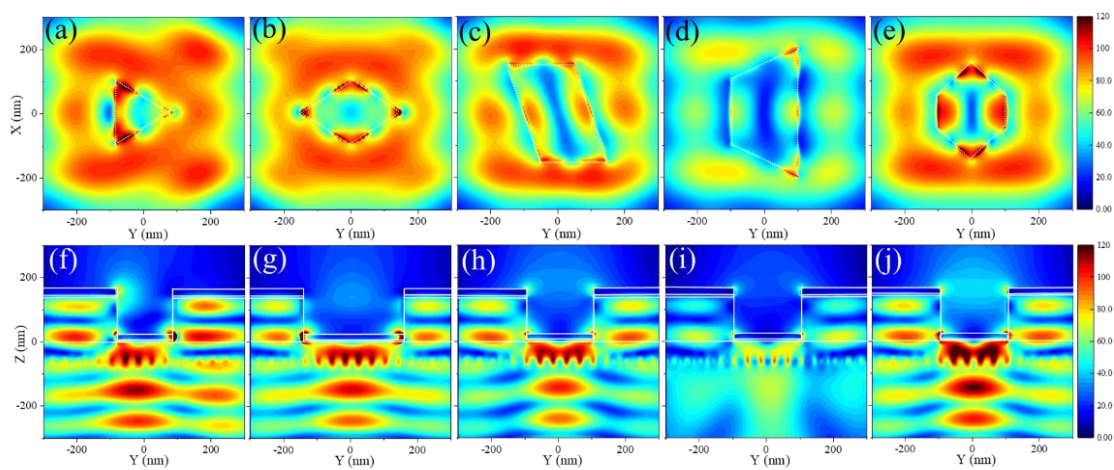

Figure 5．Simulated interaction between point dipole and the polygonal plasmonic microcavities in GaN. E-field intensity distributions for the microcavities of triangle[(a) (f)]、diamond [(b) (g)], parallelogram[(c) (h)], trapezoidal[(d) (i)] and hexagon[(e) (j)] in the X-Y and Y-Z planes, respectively.

The 440 nm radiation-emitting dipoles are positioned within the MQW region. We analyzed the E-field intensity distribution in X-Y and Y-Z planes of triangular, rhombic, parallelogram, trapezoid, and hexagonal microcavity structures (as indicated by the yellow circles in Figure 2(a)). In Figure 5, it is shown that the E-field intensity distribution at the Au/GaN interface of the five microcavities exhibited a pronounced tip effect. Specifically, triangular (Figure 5(a)), rhombic (Figure 5(b)), and hexagonal (Figure 5(e)) microcavities also displayed visible field enhancement in the inner region. The E-field distribution pattern in the X-Y plane closely follows conventional SP theory expectations, the closer to the edge of the metal microcavities, the stronger the E-field intensity [26]. But, the distribution pattern is modulated by the shape of the microcavities, with each side of the microcavities exhibiting guided mode resonance properties simultaneously[27].

However, in the Y-Z cross-section, we can see that apart from the trapezoidal microcavity structure in Figure 5(i), the other four microcavities have strong E-fields in the region under the microcavities, and the strongest E-field is not distributed in the area adjacent to the cavity bottom, **but at a certain distance.** The longitudinal distribution of this E-field is completely different from the traditional SP model, resulting in better coupling effects. The longer penetration depth results in a better overlap between the localized plasmon field and the MQW region. The hybrid effect of multiple modes greatly enhances the local electric field intensity. The shape characteristics and strong electric field localization of plasmons below the microcavity are demonstrated. This strong localized electric field will have additional effects on the localization of exciton behavior, leading to an increase in both the lifetime and recombination efficiency of excitons in the MQW.

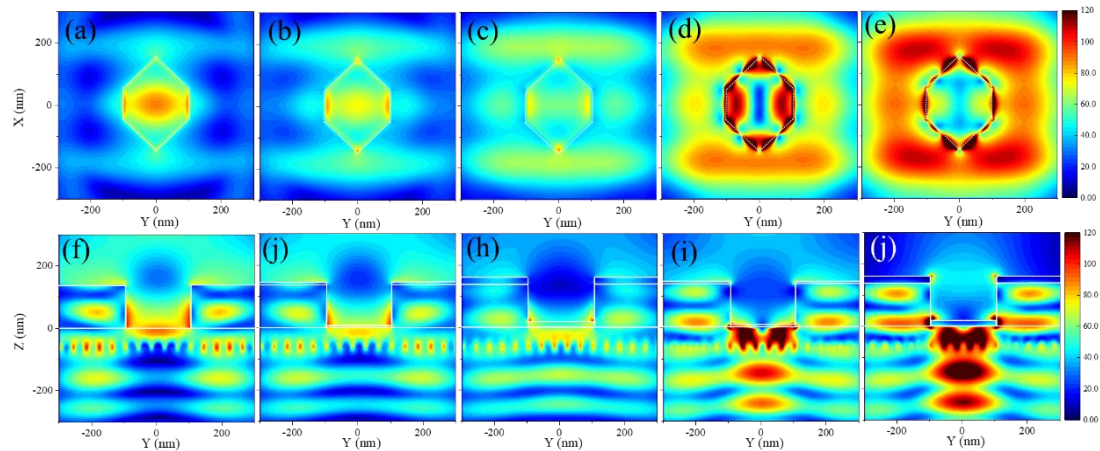

Figure 6. Cross sectional electric field distribution of the dipole source in t polygonal plasmonic microcavities. Without metal[(a) (f)]; with 5nm-thick Au film[(b) (j)]; with 20nm-thick Au film [(c)(h)]; with 5nm-thick Al film[(d)(i)]; with 20nm-thick Al film[(e)(j)].

The above demonstrates the shape effect of microcavities on plasmons, different metals may have different effects too, which is also very important. We use the hexagonal polygonal plasmonic microcavities to compute the impacts of various metals and metal thickness on the electric field intensity distribution, as shown in Figure 6.

The electric field intensity distribution for the LED with a hexagonal hole is shown in Figure 6(a), 6(f). Without any metal, the light energy density cannot extend to the outside region because the E-field energy is primarily concentrated in the hole inside. When 5 nm-thick Au film [Figure 6 (b), (J)] is applied to the hole, the E-field intensity is substantially weakened, and no much improvement with thicker 20nm-thick Au film [Figure 6 (c), (h)]. It has been inditified that Au can contribute to the SP effect at longer wavelength range not shorter than 500 nm [3], the 440 nm dipole source used in the simulation will be absorbed by Au.

However, when 5 nm-thick Al film [Figure 6 (d), (i)] is applied to the hole, the Al/GaN interface exhibits an electric hexapoles phenomenon[28]. It has also been inditified that Al can contribute to the SP effect at wide wavelength range from visible to below 300 nm [3]. An unusually strong E-field is generated around the microcavity, even with a 6-nm thick layer of Al, including the MQW region under the metal microcavity [Figure 6(i)]. When the Al thickness is 20 nm [Figure 6(e)(j)], the electric field around the metal is even stronger, and E-field intensity rise in the MQW active region. This shows a stronger microcavity coupling effect between microcavity plasmon and exciton. At the same time, corresponding to Figure 5 (e) and (j), when 5nm-thick Au and 20nm-thick Al are combined, the effect is slightly weaker than that of entire Al microcavity due to the absorption effect of Au, but this does not affect the final localized field enhancement. Therefore, these results indicate that the Cavity Plasmon effect comes from both the polygonal microcavity structure and the metal itself, making it a new type of plasmonic structure.

## 5. Conclusions

In this study, the polygonal hole structure can be created in GaN-based LEDs by utilizing nanoporous Ni as an etching mask. The polygonal metal microcavity can be realized by deposting metal film into the polygonal holes. The polygonal metal microcavities presnt noval E-field distribution patterns and plasmonic effect to the LEDs, named Cavity Plasmon effect. The light output power of the LEDs is increased by 46%. Simultaneously, the TRPL test prooves an increase in exciton lifetime. Based on the finite-difference time-domain simulation results, it is suggested that the special electric field enhancement effect from the cavity plasmon can lead to a strong localized plasmon electric field not only distributed in the X-Y plane in the immediate vicinity of the microcavity, but also forming high-intensity localized electric field hotspots in the Z direction away from the metal surface. These high-intensity localized hotspot energies provide feedback to the charge carriers in the MQW, ultimately resulting in increased carrier lifetime and improved radiative recombination efficiency. The proposed new coupling of the Cavity Plasmon effect offers a very promising option for enhancing light emission performance of the LEDs and relevant semiconductor optoelectronic devices.

**FUNDING**

This work is supported by National Natural Science Foundation of China (12074182), Collaborative

Innovation Center of Solid-State Lighting and Energy-saving Electronics.

## AUTHOR DECLARATIONS

**Conflict of Interest**

The authors have no conflicts to disclose.

**Author Contributions**

**Yuyin Li:** Software (lead); Writing-original draft (equal); Investigation (equal). **Jing Zhou:** Writing-original draft (equal); Data curation (equal). **Ziwen Yan**: Formal analysis (equal). **Xianfei Zhang**: Investigation (equal). **Zili Xie**: Methodology (equal); Project administration (equal). **Xiangqian Xiu:** Resources (equal). **Dunjun Chen**: Methodology (equal). **Bin Liu**: Methodology (equal). **Hong Zhao**: Resources (equal); **Yi Shi**: Validation (equal). **Rong Zhang**: Supervision (equal). **Youdou Zheng**: Supervision (equal). **Peng Chen**: Conceptualization (lead); Funding acquisition (lead); Project administration (lead); Writing - review & editing (lead).

## DATA AVAILABILITY

The data that support the findings of this study are available from the corresponding authors upon reasonable request.